\documentclass[a4paper]{article}
\usepackage{INTERSPEECH2022}
\usepackage{CJKutf8, cite}
\usepackage{multirow}
\usepackage{makecell}
\usepackage{color}

\newcommand\blfootnote[1]{%
\begingroup
\renewcommand\thefootnote{}\footnote{#1}%
\addtocounter{footnote}{-1}%
\endgroup
}
\title{Intermediate-layer output Regularization for Attention-based Speech Recognition with Shared Decoder}
\name{{Jicheng Zhang$^{1,3\dag}$, Yizhou Peng$^{1,3\dag}$,Haihua Xu$^2$, Yi He$^2$,  Eng Siong Chng$^3$,Hao Huang$^{1,4*}$}
\address{
    $^1$SISE, Xinjiang University, Urumqi, China \\
    $^2$AI Lab ByteDance \\
  $^3$SCSE, Nanyang Technological University, Singapore \\
  $^4$Xinjiang Provincial Key Laboratory of Multi-lingual Information Technology, Urumqi, China
  }
}
\email{}
\begin{document}
\maketitle
\begin{abstract}
Intermediate layer output (ILO) regularization  by means of multitask training on encoder side has been shown to be an effective approach to yielding improved results on a wide range of end-to-end ASR frameworks. In this paper, we propose a novel method to do ILO regularized training differently. Instead of using conventional multitask methods that entail more training overhead, we directly make the intermediate layer output as input to the decoder, that is, our decoder not only accepts the output of the final encoder layer as input, it also takes the output of the encoder ILO as input during training. With the proposed method, as both encoder and decoder are simultaneously ``regularized", the network is more sufficiently trained, consistently leading to  improved results, over the ILO-based CTC method, as well as over the original attention-based modeling method  without the proposed method employed.
\end{abstract}

\blfootnote{
    \quad  $^*$correspondence author.
}
\blfootnote{
\quad This work is done when $\dag$ are working  as interns at SCSE, MICL Lab, NTU, Singapore.
}

\section{Introduction}~\label{sec:introduction}
End-to-end (E2E) ASR modeling framework, such as recurrent neural network transducer (RNN-T)~\cite{2019Improving,2020Rnn,zhou2021language}, and attention-based encoder-decoder framework~\cite{2019Speaker,2021Recent,chorowski2015attention,zeng2018end}, say Listen, Attend and Spell (LAS)~\cite{chan2015LAS,2018Deep,yusuf2022usted},  Transformer~\cite{li2019RnnT,2020Multi,peng2021minimum}, as well as Conformer~\cite{gulati2020conformer,sun2021semantic,zeineldeen2021conformer} and etc., 
has de facto become predominant in both research and industrial areas  of speech recognition~\cite{yao2021wenet,miao2020transformer}, thanks to its simplicity, compactness, and more importantly  effectiveness in yielding improved recognition results. 

Despite much progress, E2E ASR modeling framework still faces a lot of challenges. For instance, data hungry issue is always inherent in the E2E modeling framework. This is particularly true when our models are getting deeper and deeper. 
To data, the most commonly used transformer encoder layer is 12 layers.
As a result, to maximally exploit the potentiality of such a deeper model, more data is always desired. On the other hand, 
it would be very natural for one to ask a question: given a limited data set, how should we think of an approach to releasing the potentiality of existing E2E ASR framework?  In this paper, we are meant to answer this question by means of employing an intermediate loss as a regularization term to the primary loss function.
To achieve improved E2E ASR results, extra losses, including encoder intermediate loss, being employed has been practiced for long in ASR community. For instance, 
\cite{liu2021RnnTauxiliary} has introduced a series of auxiliary loss functions to boost RNN-T performance. To let the bottom layers of the encoder have more speech content-based classification capability,  the intermediate layer output is employed as input to a shared decoder, and hence such a capability is intensified to learn. Furthermore, \cite{wang2021selfInter} also attempted to enforce the intermediate layer to have more capabilities of 
learning graphemic-state classification. Likewise, \cite{tjandra2020DejaVu} proposed to use intermediate CTC loss to boost the performance of the Transformer and Conformer encoder-based CTC ASR models. Similar to~\cite{tjandra2020DejaVu}, \cite{lee2021intermediateCtc} proposed to take advantage of intermediate CTC losses to learn context dependence for CTC ASR model.

\begin{figure} [th]
    \centering
     \includegraphics[width=8cm]{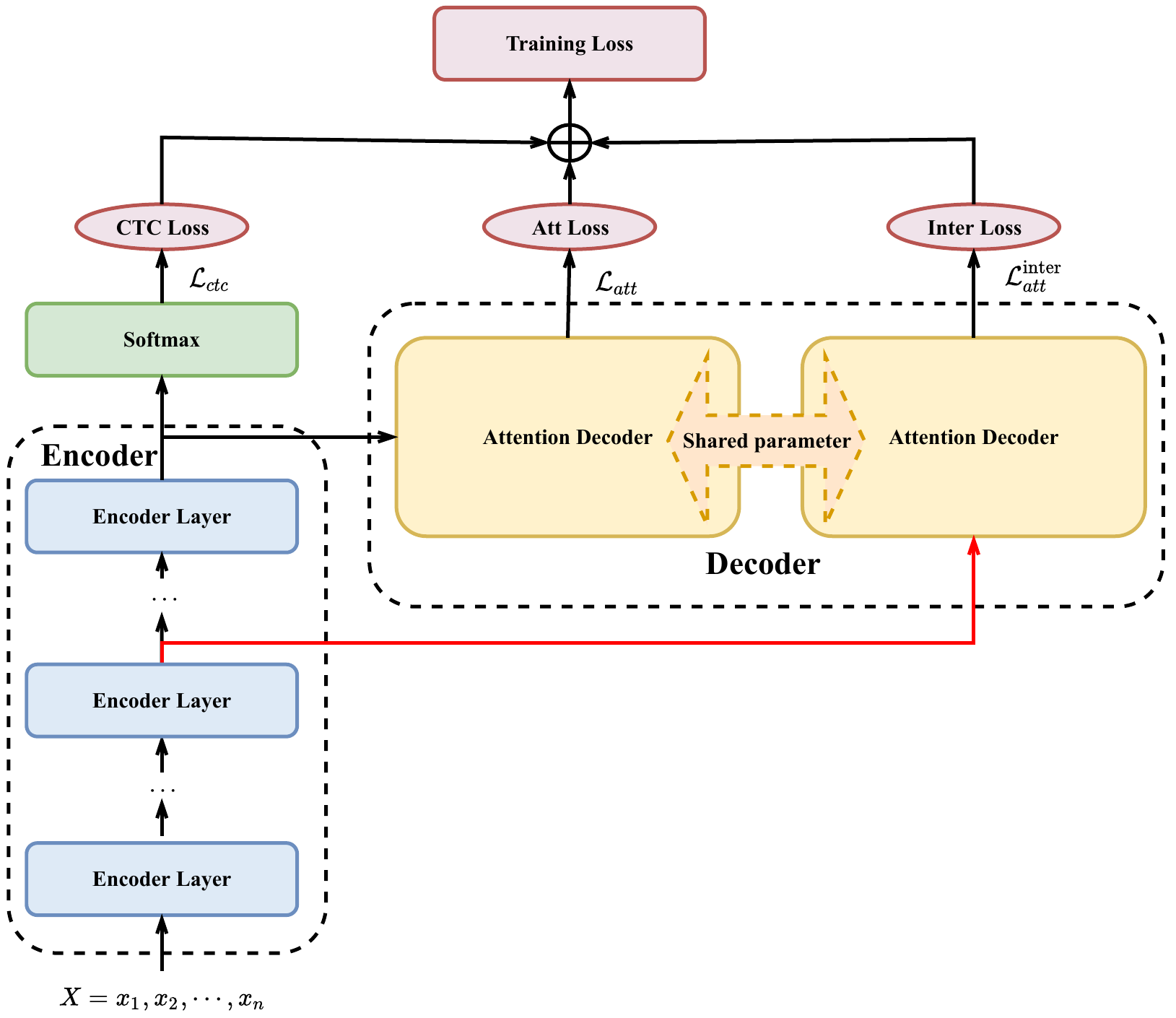}
    \caption{The architecture of the proposed intermediate layer output (ILO) regularization training method. During training, except for the normal back-propagation, both decoder and the bottom layers of encoder are also learned with an introduced $\mathcal{L}_{att}^{\text{inter}}$ loss. During decoding, the extra connection between ILO and decoder is removed, no extra computation is entailed.}
    \label{fig:network}
\end{figure}

In this paper, we propose a novel intermediate loss to boost the performance of  attention-based E2E ASR models. Different from the prior works, we let the output of the intermediate and final layers share the same decoder, as is shown in Figure~\ref{fig:network}.
The advantage of the proposed method lies in though we have auxiliary loss from encoder intermediate layer, no extra tasks are introduced. This reduces necessity of learning extra parameters. More importantly, not only is  the content-based knowledge  learning for the bottom layer of the encoder intensified, an extra input to the decoder also makes it more resilient to the change of encoder input, namely acoustic features. 

The main contribution of this paper can be summarized as follows.
1) We introduce an extra connection between  encoder intermediate layer and decoder. That is, except for taking normal input from the encoder final layer, the decoder also takes input from the encoder intermediate layer. By such a means, both encoder and decoder of our attention-based ASR system have been simultaneously regularized with such a extra back-propagation gradients, yielding consistent improved recognition results over ASR system with intermediate CTC loss as well as system without regularization at all. 2) We verified the efficacy of the proposed method over 3 public available data sets, one, an English accent data set~\cite{shi2021accented}, another, SEAME~\cite{lyu2010seame}, a Southeast Asian English-Mandarin code-switching data set, and the third, 960-hour full Librispeech data set respectively.  
3) As an ablation, we also analyze how the proposed method affects the ASR performance in detail.

\section{Related Work}~\label{sec:related}
As above-mentioned, employing encoder Intermediate Layer Output (ILO) to learn an auxiliary task  has been widely adopted for long. Generally, the prior work can be classified into two categories. One category belongs to multitask learning. For instance, \cite{2021E2E} employs ILO to learn an accent classification in addition to a normal ASR recognition task under a single E2E-based encoder-decoder framework. The other category is to employ the ILO-based auxiliary task to assist  the primary speech recognition task.
For example, 
\cite{2017Multitask} proposed to use the ILO to learn phoneme or senone classification, boosting the performance of the E2E model, which is particularly useful for deeper network. Similarly \cite{lee2021intermediateCtc} proposed intermediate losses to strengthen the learning capability of CTC models.
However, all these works have introduced an extra new task, resulting in more parameters to learn. In this paper, though we focus on ILO loss, no extra task is introduced, and hence it is much simpler.

To the best of our knowledge, the most related work is from  \cite{liu2021RnnTauxiliary}. In \cite{liu2021RnnTauxiliary}, an intermediate layer output is also taken as input to the shared decoder. However, the difference are remarkable.
First, \cite{liu2021RnnTauxiliary} thinks the intermediate layer output is more focused on feature learning and hence its directly connecting to the shared decoder is ``suboptimal". As a result, \cite{liu2021RnnTauxiliary} introduces a lightweight MLP to transform the intermediate layer output first, and then let the output of the MLP as input to the shared decoder.  In this paper, we let the intermediate output of our encoder directly connect with the decoder as a normal input instead. The proposed method is much simpler. More importantly, during back-propagation, the shared decoder for the intermediate layer is frozen for updating in \cite{liu2021RnnTauxiliary}, while  we keep updating  both encoder and decoder for the intermediate loss sequentially.
\section{Proposed Method}~\label{sec:proposed}
The proposed method is verified with the Conformer~\cite{gulati2020conformer} ASR framework.
The encoder of Conformer is  composed of blocks, which are denoted as layers in this paper. For each layer $i$, suppose the input is $x_i$, and the output is $y_i$, then the operation of each layer can be described with following equations:
\begin{equation}
    \Tilde{x}_i = x_i + \frac{1}{2}\text{FFN}(x_i)
\end{equation}
\begin{equation}
    x^{\prime}_i = \Tilde{x}_i + \text{MHSA}(\Tilde{x}_i)
\end{equation}
\begin{equation}
    x^{\prime\prime}_i = x^{\prime}_i + \text{Conv}(x^{\prime}_i)
\end{equation}
\begin{equation}
    y_i = \text{Layernorm}(x^{\prime\prime}_i) + \frac{1}{2}\text{FFN}(x^{\prime\prime}_i)
\end{equation}
where FFN, MHSA, Conv, and Layernorm refer to feed-forward network, multi-head self-attention,  convolution, and layer normalization~\cite{layer-normalization} operations respectively. 
As a result, the so-called intermediate output refers to  some $y_i$, where $1\leq i < N$, and N is the total layers of Conformer encoder.

With the Conformer framework, the baseline ASR system is trained with multitask learning method, that is, our encoder is optimized with CTC loss~ $\mathcal{L}_{ctc}$, while the entire network is optimized with ASR negative log posterior, i.e., $\mathcal{L}_{att}$.   Combining the two losses, the overall training loss is 
\begin{equation}
  \mathcal{L}_{asr}={\alpha\mathcal{L}}_{ctc}+{(1-\alpha)\mathcal{L}}_{att}
  \label{eqn:baseline-loss}
\end{equation}
where $\alpha$ is a weighing factor, and it is fixed with 0.3 in what follows.  In~\cite{Watanabehybrid,xiao2018hybrid}, it is shown employing  such a combined loss of Eqn. \ref{eqn:baseline-loss} can consistently yield improved recognition results.

For the proposed method as illustrated in Figure~\ref{fig:network},we employ the encoder intermediate output as an extra input to the decoder. Consequently, we introduce another loss, denoted as $\mathcal{L}_{att}^{\text{inter}}$, into Eqn~\ref{eqn:baseline-loss}, yielding the new loss:
\begin{equation}
    \mathcal{L}_{asr}^{\prime}= \alpha\mathcal{L}_{ctc} +
    \beta \mathcal{L}_{att} +
    \gamma\mathcal{L}_{att}^{\text {inter}}
    \label{eqn:proposed-loss}
\end{equation}
where we let $\alpha+\beta+\gamma=1$. As the proposed method is very close to an ILO-based multitask learning method, that is, we let the encoder ILO connect with another recognition or classification task, such as a CTC-based ASR task, and etc, it is necessary to compare the proposed method with such an ILO-based multitask method, particularly CTC method. For completeness, the loss of the ILO-based CTC method is rewritten as follows:  
\begin{equation}
    \mathcal{L}_{asr}^{\prime\prime} =  \alpha\mathcal{L}_{ctc} +
    \beta \mathcal{L}_{att} +
    \gamma \mathcal{L}_{ctc}^{\text{inter}}
    \label{eqn:inter-ctc-loss}
\end{equation}
where we also restrict $\alpha+\beta+\gamma=1$.

\begin{figure} [th]
    \centering
     \includegraphics[width=8cm]{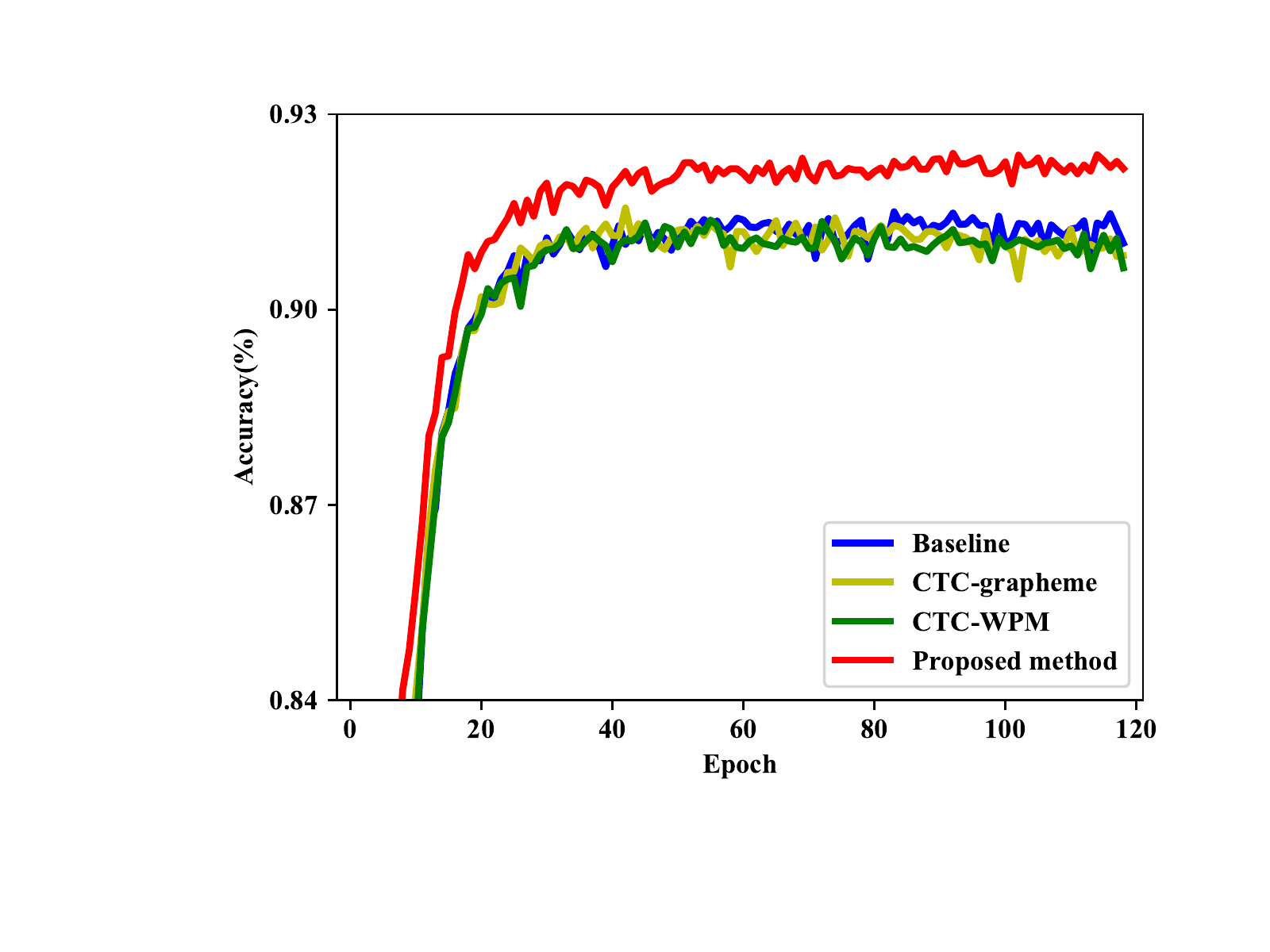}
    \caption{Validation accuracy versus training epoch for the three ASR systems trained with losses of Equations~\ref{eqn:baseline-loss}, \ref{eqn:proposed-loss} and \ref{eqn:inter-ctc-loss} respectively. The curve is drawn with the accented English data  set, as detailed in Section~\ref{sub:exp:data}}
    \label{fig:acc}
\end{figure}

From Equations~\ref{eqn:baseline-loss}, \ref{eqn:proposed-loss} and \ref{eqn:inter-ctc-loss}, we can produce three ASR systems with three losses to train corresponding models.
To check the efficacy of the proposed method simply, we can compare the accuracy on the validation data  between the three ASR systems during training process.
Figure~\ref{fig:acc} plots the validation accuracy versus training epochs for the three methods. 
For CTC-grapheme and CTC-WPM, it means that the middle layer sends grapheme based CTC and subword based CTC respectively.

From Figure~\ref{fig:acc}, we notice that the proposed method obtains consistently best accuracy on the validation data during training. 
The CTC-grapheme and CTC-WPM method has not shown clear improvement over the baseline ASR system on the accented English data set.
Further performance results on different test sets will be presented in 
Section~\ref{sec:exp}.

It is worth noting that once we use the ILO regularization to train the network, we disconnet the connection between ILO and decoder. That is, our decoding network is completely the same with the baseline without ILO employed at all. However, our decoder is still a multitask based one, i.e., the decoding output is a combination of CTC and attention output like what is proposed in~\cite{xiao2018hybrid}.


\section{Experiment}~\label{sec:exp}
\vspace{-8mm}
\subsection{Data}~\label{sub:exp:data}
To verify the efficacy of the proposed method, we conduct experiments on three publicly available data corpora.
They are Accented English data corpus~\cite{shi2021accented}, 960 hours of Librispeech data, as well as SEAME data~\cite{lyu2010seame}, a Southeast Asia English-Mandarin code-switching data set.
Table~\ref{tab:data-spec} summarizes the details of the overall data sets that are employed to evaluate the proposed method in what follows.

The Accented English data corpus is released by \texttt{DataTang}, in an Accented English Speech Recognition Challenge \& Workshop of 2019. The purpose is targeted for both accent and speech recognition challenges.
It's a read speech and the total 8 accents are from  American(US), British(UK), Chinese(CHN), Indian(IND), Japanese(JPN), Korean(KR), Portuguese(PT) and Russian(RU) accent respectively. Each accent has roughly 20 hours. One can refers to~\cite{shi2021accented} for more details.

For the SEAME code-switching data set, the total length of the data is about 110 hours. It is a spontaneous conversational speech corpus, recorded under clean environment.
To evaluate the performance, two code-switching test sets are defined. One test set is denoted as Test\textsubscript{man}, of which
the speech content is dominated with Mandarin, while another test set is named as Test\textsubscript{sge} that is 
biased to Southeast Asian English speech, of which the speakers are mainly from Malaysia and Singapore. For more details, one can refer to~\cite{lyu2010seame}.
\subsection{Models}~\label{sub:exp:model}
All experiments are performed using E2E-based Conformer
modeling framework with Espnet tookit~\cite{watanabe2018espnet}.
We use the acoustic feature that
is 83-dimensional with 80-d filter-bank plus 3-d pitch features~\cite{ghahremani2014pitch}.
In all experiments, we fix the Conformer with 12-layer  encoder and 6-layer decoder.
For the Accented English and SEAME  experiments, both the encoder and decoder with 4 attention heads, the attention dimension is
256.
But for Librispeech experiments, the attention heads are increased to 8 and corresponding attention dimension increased to 512 dimensions accordingly.

During training, we proceed with 0.1 dropout~\cite{2014Dropout}, and our Conformers are optimized with Noam optimizer~\cite{vaswani2017attention}. 
For English data, we employ Byte Pair Encoding(BPE)~\cite{sennrich2015neural} to
train Word Piece Model (WPM) from corresponding 
English transcriptions.
For SEAME code-switching dataset, in order to better balance the number of Chinese and English modeling units, 
we let the English data
have 2920 BPEs that are  comparable with 2600  single Chinese characters from the entire training set. 
For Accented English and Librispeech datasets, the BPE sizes are 2000 and 5000 respectively.
To yield state-of-the-art performance, we employ 
SpecAugment~\cite{INTERSPEECH-SpecAug-2019} for all models. However, only SEAME data has been applied with speed perturbation~\cite{ko-2015-speed_perturb}.

For inference, all the final models are obtained by averaging the model parameters of the best 5 epochs.
We fix the beam size with 10. Besides, to yield better results, the final posterior is interpolated with  CTC and attention-based decoder output, and the scaling factors are 0.2 versus 0.8 respectively. We haven't tweeted other parameters specifically.
\begin{table}[htbp]
 \centering
 \caption{Overall Data description}
 \label{tab:data-spec}
\begin{tabular}{c c c c c}
 \toprule
   Corpus &Set & Len (Hrs) & Word/Utt. & Sec./Utt. \\
   \midrule
   \multirow{3}*{\makecell{Accented \\ English}} 
   & Train & 148.5 & 9.72 & 4.29\\
   & Dev& 14.5 & 9.66 & 4.35 \\
   & Test& 20.95 & 9 & 4.15 \\
   \midrule
   \multirow{3}{*}{Librispeech} & Train & 961 & 33.43 & 12.3\\
     & Test\textsubscript{clean}& 5.4& 20.0 & 7.4\\
     & Test\textsubscript{other} & 5.3& 17.8& 6.5 \\
    \midrule
   \multirow{3}*{\makecell{SEAME}} 
     & Train & 93.6 & 14.14 & 3.9 \\
     & Test\textsubscript{man} & 7.5 & 14.76 & 4.1 \\
     & Test\textsubscript{sge} & 3.9 & 10.12 & 2.7 \\
    \bottomrule
\end{tabular}
\end{table}
\section{Results}~\label{sec:res}
Table~\ref{tab:accent-res01} presents the ASR performance of the proposed method on the accented English data set. From Table~\ref{tab:accent-res01}, we notice that the proposed method (M3) has achieved 8.2\% relative WERR Word Error Rate Reduction(WERRR) over the baseline Conformer model (M1)  while it (M4) has gained 4.9\% relative WERR over the model with SpecAug-based data augmentation~(M2) on test set respectively. 
After using SpecAug, we found that the effectiveness of the proposed method decreased, but the performance is still significantly improved.
In other words, our proposed method can be effectively combined with SpecAug method to produce better results.

More interestingly, system M5 obtains the best WER once we combine the proposed method with the vanilla ILO-based CTC method using grapheme as output. Specifically, M5 obtains 8.6\% relative WERR over M2 system. 
Finally, ILO-based CTC-grapheme and ILO-based CTC-WPM methods both obtain smaller WERR over baseline system compared with our proposed method.
\begin{table}[htbp] 
    \centering
        \caption{WERs(\%) for the Accented English data, where ILO-CTC-g refers to ILO-based CTC with grapheme as output units while ILO-CTC-WPM stands for ILO-based CTC with WPM as output.}
    \begin{tabular}{c l c   c c}
        \toprule
        \multirow{2}*{ID} & \multirow{2}*{Model} & \multirow{2}*{(Layer,$\gamma$)}& \multicolumn{2}{c}{WER(\%)}  \\ \cline{4-5}
        & & & Dev & Test \\
        \midrule
        \multirow{1}*{M1}
            & Conformer & -  & 9.5 & 11 \\
        \multirow{1}*{M2}
            & M1 + SpecAug & -  & 6.9 & 8.1 \\
        \multirow{1}*{M3}
            & Proposed & (9,0.2) & 8.8 & 10.1 \\
        \multirow{1}*{M4}
            & M3+SpecAug &  (9,0.2)  & 6.7 & 7.7 \\
        \multirow{1}*{M5}
            & M4 + ILO-CTC-g &  (9,0.2)  & 6.4 & 7.4 \\
        \multirow{1}*{M6}
            & ILO-CTC-g &  (9,0.15)  & 9.3 & 10.7 \\
        \multirow{1}*{M7} 
            & M6 + SpecAug & (9,0.15)  & 6.8 & 8 \\
        \multirow{1}*{M8} 
            & ILO-CTC-WPM & (9,0.1)  & 9.4 & 10.7 \\
        \multirow{1}*{M9} 
            & M8 + SpecAug & (9,0.1)  & 6.9 & 8 \\
        \bottomrule
    \end{tabular}
    \label{tab:accent-res01}
\end{table}
Table~\ref{tab:libri-res} reports the WER performance of the proposed method on Librispeech data set. We can see from Table~\ref{tab:libri-res} the proposed method is more effective when no SpecAug is employed, the proposed method(M3) can get $\sim$8.2\% relative WERR over the baseline system (M1) on Test\textsubscript{other}.
However, when using SepcAug, our method(M4) only gets 1.4\% relative WERR  over M2 on Test\textsubscript{other}.
This again suggests the proposed method
might be more beneficial for low-resource ASR modeling. Due to computing resource limitation, we omit the experiments for ILO-CTC-grapheme, as well as ILO-CTC-WPM.
\begin{table}[htbp] 
    \centering
        \caption{WERs (\%) for Librispeech data.}
    \begin{tabular}{c l c  c  c c}
        \toprule
        \multirow{2}*{ID} & \multirow{2}*{Model} & \multirow{2}*{(Layer,$\gamma$)}  & \multicolumn{2}{c}{WER(\%)}  \\ \cline{4-5}
        & & & Test\textsubscript{clean} & Test\textsubscript{other} \\
        \midrule
        \multirow{1}*{M1} 
            & Conformer &  -  & 3.4 & 9.8 \\
        \multirow{1}*{M2} 
            & M1 + SpecAug & -  & 2.9 & 6.9 \\
        \multirow{1}*{M3} 
            & Proposed  & (9,0.2) & 3.2 & 9.0 \\
        \multirow{1}*{M4} 
            & M3 + SpecAug & (9,0.2) & 2.8 & 6.8 \\
        \bottomrule
    \end{tabular}
    \label{tab:libri-res}
\end{table}

Table~\ref{tab:seame-res} presents the WER performance of the proposed method on SEAME data set. 
We notice that the proposed method has achieved consistent performance improvements on both Test\textsubscript{man} and Test\textsubscript{sge} test sets with or without SpecAug.
However, both ILO-based CTC-grapheme and ILO-based CTC-WPM methods gain no improved performance compared with our proposed method on both test sets(M5,M7 versus M3 and M6,M8 versus M4).

\begin{table}[htbp] 
    \centering
        \caption{WERs(\%) for Seame data sets}
    \begin{tabular}{c l c  c c}
        \toprule
        \multirow{2}*{ID} & \multirow{2}*{Model} & \multirow{2}*{(Layer,$\gamma$)} & \multicolumn{2}{c}{WER(\%)}  \\ \cline{4-5}
        & & & Test\textsubscript{man} & Test\textsubscript{sge} \\
        \midrule
        \multirow{1}*{M1}
            & Conformer &  - & 18.51 & 25.9 \\
        \multirow{1}*{M2}
            & M1 + SpecAug &  - & 16.38 & 23.28 \\
        \multirow{1}*{M3}
            & Proposed & (9,0.2) & 18.42 & 25.48 \\
        \multirow{1}*{M4} 
            & M3 + SpecAug & (9,0.2)  & 15.97 & 22.82 \\
        \multirow{1}*{M5}
            & ILO-CTC-g & (9,0.15) & 18.76 & 25.65 \\
        \multirow{1}*{M6}
            & M5 + SpecAug & (9,0.15) & 16.25& 22.91 \\
        \multirow{1}*{M7} 
            & ILO-CTC-WPM & (9,0.1)  & 18.86 & 26.47 \\
        \multirow{1}*{M8} 
            & M7 + SpecAug & (9,0.1)  & 16.49 & 23.16 \\
        \bottomrule

    \end{tabular}
    \label{tab:seame-res}
\end{table}

\section{Analysis}~\label{sec:analysis}
\vspace{-8mm}
\subsection{Layer of ILO Effect}~\label{sub:analysis:layer}
\begin{figure} [ht]
    \centering
     \includegraphics[width=7cm]{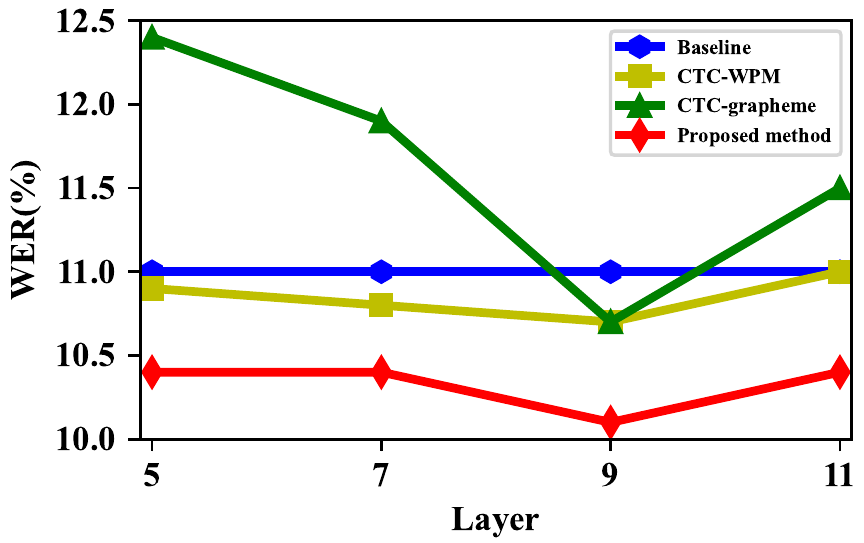}
    \caption{WER (\%) versus specific intermediate layer output on Accented English test set. 
    The Conformer encoder has 12 layers in total.
    }
    \label{fig:multi-units-wer-versus-layer}
\end{figure}

The proposed ILO-based CTC-WPM and ILO-based CTC-grapheme methods are all employing intermediate layer output to regularize the training. Here, we want to know which intermediate layer can yield optimal results. Figure~\ref{fig:multi-units-wer-versus-layer} plots WER versus different intermediate layer output on the Accented English data sets for the three methods. We can see that all of the three methods achieved best WERs around layer 9 with the 12-layer encoder setup.
\subsection{Effect of the proposed ILO regularization}
From Tables~\ref{tab:accent-res01} to \ref{tab:seame-res}, we can see the proposed ILO regularization method has achieved consistent performance improvement over the baseline and other ILO-based multitask methods.
Naturally, we are curious about to what extend the proposed method affects both encoder and decoder. To this end, we let the Conformer that is trained with the proposed method decode with different mode. Specifically, 
we let the encoder final layer CTC decode, and see how the encoder is affected.
Though we cannot easily verify how the decoder alone is affected, we can see how pure attention-based decoder results are affected.
Table~\ref{tab:ilor-ana} reports the ``decomposed" WERs in each case. We can see from Table~\ref{tab:ilor-ana}, the most affected part is the encoder as the CTC system gets better performance improvement, i.e., 6.8\% 
versus 3.66\% relative WERR
when M4 is compared with M2 in ``CTC"  versus  ``Attention" case respectively.
\begin{table}[htbp] 
    \centering
        \caption{ WERs(\%) of using the proposed ILO regularization method, where ``CTC" refers to using pure CTC output to infer,  ``Attention" refers to  using pure attention-based decodder output to infer, and while ``Hybrid" is the interpolated inference method, 
        of which the CTC and attention-based decoder output factors are 0.2 and 0.8 respectively
        }
    \begin{tabular}{c l c c c}
        \midrule
        \multirow{2}*{ID} & \multirow{2}*{Model} & \multicolumn{3}{c}{Test set(WER\%)}  \\ \cline{3-5}
        & & CTC & Attention & Hybrid \\
        \hline
        \multirow{1}*{M1}
            & Conformer  & 13.4 & 11.3 &11 \\
        \multirow{1}*{M2}
            & M1 + SpecAug  & 11.7 & 8.2 &8.1 \\
        \multirow{1}*{M3}
            & Proposed & 12.5 & 10.4 & 10.1\\
        \multirow{1}*{M4}
            & M3 + SpecAug & 10.9 & 7.9 & 7.7\\
        \bottomrule
    \end{tabular}
    \label{tab:ilor-ana}
\end{table}
\section{Conclusions}
In this paper, we proposed an intermediate layer output regularization method to train state-of-the-art Conformer model for speech recognition. Different from the prior works, the regularization is realized with an extra connection between the intermediate layer  of encoder and the decoder. Consequently, the proposed method is much cheaper to train. We verified the efficacy of the proposed method on three different publicly available data sets. It has achieved consistent WER reductions, when it is compared with the baseline Conformer, as well as various intermediate layer output CTC-loss-based Conformer.
	\vfill\pagebreak
\clearpage
\bibliographystyle{IEEEtran}

\bibliography{mybib}

\end{document}